\documentclass{article}

\usepackage[final]{drlw_neurips_2019}

\usepackage[utf8]{inputenc} 
\usepackage[T1]{fontenc}    
\usepackage{hyperref}       
\usepackage{url}            
\usepackage{booktabs}       
\usepackage{amsfonts}       
\usepackage{nicefrac}       
\usepackage{microtype}      
\usepackage{graphicx}
\usepackage{enumitem}
\usepackage{array,multirow}
\usepackage{media9}
\usepackage{hyperref}
\usepackage{amsmath,amssymb}

\newcommand\norm[1]{\left\lVert#1\right\rVert}

\hypersetup{
    colorlinks=true,
    linkcolor=black,
    filecolor=black,      
    citecolor=black,
    urlcolor=cyan
}

\title{Reinforcement Learning for High-dimensional Continuous Control in Biomechanics: An Intro to ArtiSynth-RL}

\author{%
  David S.~Hippocampus\thanks{Use footnote for providing further information
    about author (webpage, alternative address)---\emph{not} for acknowledging
    funding agencies.} \\
  Department of Computer Science\\
  Cranberry-Lemon University\\
  Pittsburgh, PA 15213 \\
  \texttt{hippo@cs.cranberry-lemon.edu} \\
}

\author{
Amir H. Abdi \And
Masoud Malakoutian \And
Thomas Oxland \And
Sidney Fels \AND \\
University of British Columbia, Vancouver, BC, Canada\\
\\
\texttt{
\{amirabdi, ssfels\}@ece.ubc.ca,
masoudm@alumni.ubc.ca,
toxland@mail.ubc.ca
}
}

\begin{document}

\maketitle

\begin{abstract}
Neural control is an exciting  mystery which we instinctively master. Yet, researchers have a hard time explaining the motor control trajectories.
Physiologically accurate biomechanical simulations can, to some extent, mimic live subjects and help us form evidence-based hypotheses. 
In these simulated environments, muscle excitations are typically calculated through inverse dynamic optimizations which do not possess a closed-form solution.  
Thus, computationally expensive, and occasionally unstable, iterative numerical solvers are the only widely utilized solution.
In this work, we introduce ArtiSynth, a 3D modeling platform that supports the combined simulation of multi-body and finite element models, and extended to support reinforcement learning (RL) training.
we further use ArtiSynth to investigate whether a deep RL policy can be trained to drive the motor control of a physiologically accurate  biomechanical model in a large continuous action space.
We run a comprehensive evaluation of its performance and compare the results with the forward dynamics assisted tracking with a quadratic objective function.
We assess the two approaches in terms of correctness, stability, energy-efficiency, and temporal consistency.
\end{abstract}

\section{Introduction}

Motor control is a set of time-varying actuator activations generating motion in mechanical systems.
In biomechanics, the main active muscle property which varies across muscles, fluctuates during locomotion, and drives other active and passive muscle properties, is the muscle activation level.
Knowledge of the muscle activation trajectories in the human body during daily activities are crucial for various applications from  muscle driven computer animation and imitation learning to ergonomics, rehabilitation, and surgical planning. 
Muscle activations reflect the underlying neural excitations. Neural activities of shallow muscles can be, to some extent, captured directly (\textit{in vivo}) via electromyography (EMG), based on which muscle activations are indirectly estimated.
However, there are many concerns regarding the invasiveness, sensitivity, and reproducibility of this modality of data~\cite{Vigotsky2018}. 
Moreover, the inaccessibility of the hundreds of muscles in the human body, as well as the unknown relationship between the EMG and the neural excitations, degrade the value of such empirical studies~\cite{al2017analysis}.
Estimation of muscle activations, among many other unknown biomechanical variables, have led to the expansion of computational musculoskeletal models which aim to understand the neural and motion patterns of humans among other species.


The muscle forces  are  functions of their mechanical properties and neural excitations.
In computational simulations, muscle forces are estimated from kinematics via inverse-dynamics analysis. This approach solves a static optimization problem for each timestep to find the most likely set of muscle activations which will drive the model closer to the desired position~\cite{otten2003}.
Although this method is frequently utilized due to its low computational overhead~\cite{WINBY2009}, it is quite sensitive to the given motion trajectories. 
Moreover, given the static paradigm of the solution, there is no guarantee that the forward dynamics of the estimated excitations would result in a smooth motion trajectory.

An extension to the static optimization is the forward-dynamics assisted tracking (FDAT) where consecutive actions are considered  to impose temporal consistency.
This also allows for the inclusion of muscle contraction dynamics as a regularization factor to reduce sensitivity to the input kinematics~\cite{erdemir2007model}.
The cost function can also include a subset of joints' loads, tracking error, or any other time-varying variables of interest.
Regardless, the iterative quadratic solvers used in the FDAT solutions are computationally expensive. 
Other alternatives, where the cost function is minimized for a complete episode of movement, are intriguing  but impractical for real-time motion synthesis and computationally infeasible high-order inverse problems~\cite{anderson2001static}. 


Reinforcement learning (RL) is an adaptive control strategy which scales well to high-dimensional state and action spaces.
Therefore, it is getting traction in motor control and motion planning in robotics and biomechanical models.
The RL paradigm is very similar to the feedback and circuit reduction processes used by the basal ganglia of the brain during habit development and it is suggested that understanding the RL-based control mechanism is crucial to the analysis of human movement~\cite{Yin2006}.
Recent interdisciplinary collaborations have helped bridge the advances of reinforcement learning to biomechanical models~\cite{Kidzinski2018}. 

Deep RL solutions are demonstrating ground-breaking results for discrete and continuous control problems. One of the recent intriguing applications of deep RL is the continuous control of physics-based models~\cite{Lillicrap2015,Schulmanetal_ICLR2016}.
In the continuous action space, policy gradient methods such as DDPG, TD3 and SAC are showing promise~\cite{fujimoto2018addressing,haarnoja2018soft,Lillicrap2015}. 
However, the sample inefficiency of policy gradient methods is a remaining concern.

With the above motivations in mind, this paper investigates the applicability of reinforcement learning to muscle excitation estimation in a high-dimensional biomechanical model with more than 200 actuators.
We evaluate the trained policy comprehensively  against a well-established numerical solver in terms of stability, temporal consistency, energy efficiency, and correctness.
In this work, we also highlight an issue with the clipped surrogate formulation of the proximal policy optimization algorithm which can result in unreasonable policy updates. Consequently, we propose a trick for the objective function to make the training safer. 
As our last contribution,  
we have transformed  the open-source mechanical simulation platform of ArtiSynth into an RL-friendly environment.
ArtiSynth is recognized as a rich physics-based platform and we are hoping to join forces with the computer science community by making its tools and models available for RL research.

\section{Methods}

\subsection{Motion Control With Reinforcement Learning}

Proximal policy optimization (PPO) algorithms approximately enforce the Kullback–Leibler (KL) constraint, \emph{i.e.} trust regions, as a penalty in the objective function without computing the natural gradients~\cite{schulman2017proximal}.
As a result, they strike a balance between ease of implementation, sample efficiency, and ease of tuning compared to other policy gradient methods such as trust region policy optimization~\cite{TRPO}. 

The  motor control solution investigated in this paper consists of an internal forward-dynamics musculoskeletal model (described in Section~\ref{sec:musc_model}) driven by a parameterized  policy trained using the PPO algorithm. 
In our experiments, we leveraged  the clipped surrogate version of PPO which approximately maximizes the objective function while restricting the updates to smaller values.
This objective function can be summarized as
\begin{equation}
\label{eq:ppo}
L^{clip}(\theta)=  \hat{\mathbb{E}}[min(r(\theta)\hat{A_{\theta}}, clip(r(\theta), 1-\epsilon, 1+\epsilon)\hat{A_{\theta}})],
\end{equation}
where $\hat{A_{\theta}}$ is an estimator of the advantage function at the current timestep and $r(\theta)=\frac{\hat{\pi}_\theta(a|s)} {\pi_{\theta}(a|s)}$
is the probability ratio of the new policy over the old policy.

When the parameterized policy is represented by a neural network
whose weights are shared with the value function network, 
the surrogate objective in Eq.~\ref{eq:ppo} can be  augmented by adding the squared-error loss term, $L^{VF}(\theta)=(V_{\theta}(s)-V^*)^2$.
However, if the two networks don't share any learnable parameters, as is in our case, the loss terms can be kept separate.

To encourage exploration, an entropy term can be included in the objective function, as follows
\begin{equation}
\label{eq:finalppo}
L(\theta)=  \hat{\mathbb{E}}[L^{clip}(\theta)+c_{s}S_{\pi_\theta}(s)].
\end{equation}
The entropy depends on the action space distribution.
Assuming the actions to be sampled from a multi-dimensional Gaussian distribution during training, whose learnable parameters are included in $\theta$, the entropy term can be calculated in each timestep as
\begin{equation}
\label{eq:entropy}
S_{\pi_\theta}(s) = 0.5(1+ln(2\pi) + ln(\delta)),
\end{equation}
where $\delta$ is the standard deviation of the distribution.

\subsubsection{Stabilizing Trick}
Since the probability ratio, $r(\theta)$, is
lower-bounded by zero, 
the parameter updates of the  clipped surrogate PPO formulation 
is constrained when the advantage ($\hat{A}_\theta$) is positive. 
However, when the advantage of a state-action pair is negative, Eq.~\ref{eq:ppo} is reduced to 
\begin{equation}
    \label{eq:trick}
L^{clip}(\theta, \textbf{s},\textbf{a})=  max(r(\theta), 1-\epsilon)\hat{A_{\theta}}(\textbf{s},\textbf{a}).
\end{equation}
In this case, when the probability of the action in the new policy is much larger than the old policy
\big($\hat{\pi}_\theta(\textbf{a}|\textbf{s}) \gg \pi_\theta(\textbf{a}|\textbf{s})$\big),
the probability ratio $r(\theta)$, 
and consequently the objective function, $L^{clip}$, would have extremely large values which destabilises the training.
To avoid this, we slightly changed Eq.~\ref{eq:trick} and  imposed an upper bound, parameterized by $\beta$, as
\begin{equation}
    \label{eq:trick_fixed}
L^{clip}(\theta, \textbf{s},\textbf{a})=  
clip(r(\theta), 1-\epsilon,1+ \beta)
\hat{A_{\theta}}(\textbf{s},\textbf{a}).
\end{equation}

\subsubsection{Reward Function}
In our experiments, the primary goal is to find an accurate (in terms of reaching the target) and energy-efficient (in terms of minimum muscle activation) solution to the inverse-dynamics problem. 
Moreover, the motion is desired to be smooth.
Accordingly, the RL reward function is formulated as 
\begin{equation}
\label{eq:reward}
R(\textbf{s}_t,\textbf{a}_t)=
\Gamma_\delta(\textbf{u}_t, \textbf{u}^*) -
w_u \phi_u - 
w_d \phi_d - 
w_r \phi_r, 
\end{equation}
where $\textbf{u}_t$ and $\textbf{a}_t$ represent the state and activations at time step $t$, respectively.
Here,
$\Gamma_\delta(.)$
denotes a  bonus reward given to the agent when its distance to  the target state ($\textbf{u}^*$) is below the pre-defined threshold, $\delta$.

The three remaining terms in Eq.~\ref{eq:reward} are penalties based on distance to desired state, fluctuations in muscle activations and excessive activations.
At every time step, the agent is penalized based on its distance to the target state according to the $\phi_u$ term.
The third term ($\phi_d$) is to encourage temporal consistency (similar to damping) in consecutive time steps, while the fourth one is a regularization term to minimize the magnitude of muscle activations (\emph{i.e.} neural excitations).
The details of each term is discussed in the next subsection.




\subsection{Motor Control With Forward Dynamics Assisted Tracking}

To bring the RL solution into an interpretable perspective, we implemented a Forward Dynamics Assisted Tracking (FDAT) method which aims for the same set of objectives as those discussed in the RL approach.
Similar to the RL paradigm, in the FDAT method, motion terms are formulated as \emph{goals} for the dynamic model  rather than explicitly prescribed objectives.
For the methods to be relatively comparable, we designed the convex cost function of  Eq.~\ref{eq_total_qp} similar to the reward function of Eq.~\ref{eq:reward},  with three quadratic terms~\cite{stavness2010byte},
\begin{equation}
\label{eq_total_qp}
\phi_{total} = w_u \phi_u +  w_d \phi_d + w_r \phi_r.    
\end{equation}
This inverse-dynamics optimization is solved at each timestep to find the most likely muscle activations for the forward dynamics motion. 
We form a linear complementarity problem (LCP) and solve it using the quadratic  Cottle-Dantzig's approach known as the principal pivoting method (PPM).
At each timestep, this solver iteratively minimizes the cost function, $\phi_{total}$,  to best reproduce the assigned trajectories while satisfying other applied constraints. 

In Eq.~\ref{eq_total_qp}, $\phi_u$ is the kinematic term defined as the squared Euclidean distance of the desired position to the current position, \emph{i.e.},
\begin{equation}
\phi_{u} = \frac{1}{2 \Delta t} 
\norm{ \textbf{u}-\textbf{u}^*}^2.
\end{equation}
This term is normalized by the hyper-parameter $\Delta t$, \emph{i.e.}, the expected time to reach the target, which gives it a notion of velocity.
The other two terms, $\phi_r$ and $\phi_d$, tackle the muscle redundancy of the model. 
Temporal consistency is encouraged by the damping term as 
\begin{equation}
    \phi_d = \frac{1}{2} \norm{\textbf{a}_t-\textbf{a}_{t-1}}^2,
\end{equation}
which enforces smooth trajectories by minimizing the differences between consecutive muscle activations $\textbf{a}_t$ and $\textbf{a}_{t-1}$.
Lastly, the regularization component, $\phi_r$, minimizes the squared $l2$-norm of the activation vector,
\begin{equation}
    \phi_r = \frac{1}{2} \norm{\textbf{a}}^2.
\end{equation}
Therefore, increasing the coefficient, $w_r$, would result in less activated muscles which is usually desirable in optimal control problems. 
Other regularization terms may also be used such as the $l1$-norm to encourage activation sparsity.

\subsection{ArtiSynth-RL Platform}

\begin{figure}[t]
\centering
\includegraphics[width=1.0\textwidth, trim=8 8 8 8, clip]{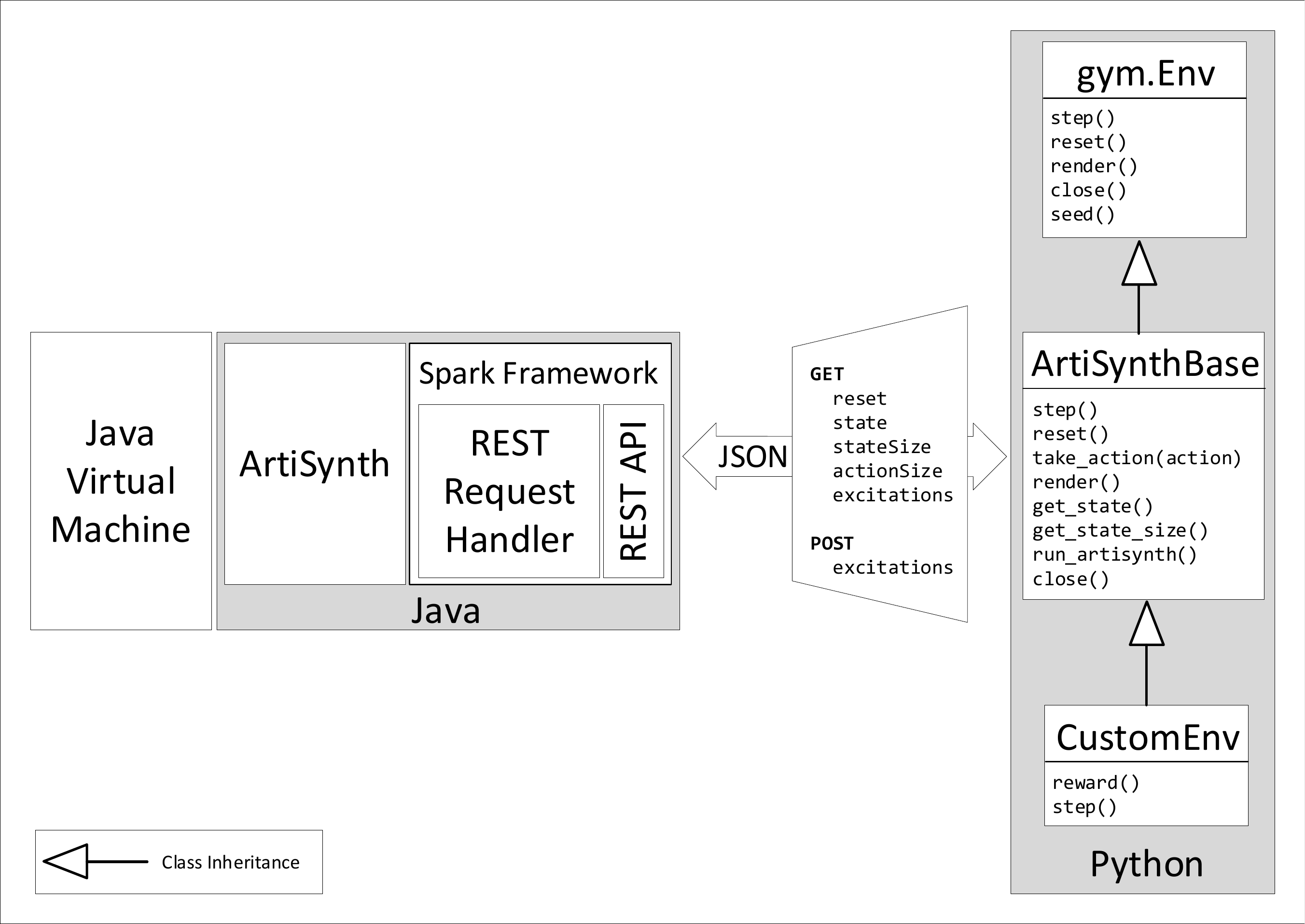}
\caption{The ArtiSynth modeling tool is extended with a RESTful api towards a platform-agnostic scalable RL-friendly  service, referred to as the ArtiSynth-RL. 
The class diagram of a Gym-compatible solution in Python is depicted on the right as a framework to guide further developments. The ArtiSynth-RL plugin and sample  Gym-compatible environments are available  \href{https://github.com/amir-abdi/artisynth-rl}{here}.}
\label{fig:artisynth-rl}
\end{figure}

ArtiSynth is an open-source physics-based mechanical modeling toolkit which combines finite elements with multi-body dynamics and solves them in a fast and convenient fashion~\cite{Lloyd2012}.
ArtiSynth is widely used in the biomedical and biomechanical research, and extendable to  general purpose mechanical and robotics simulations  (\href{www.artisynth.org}{www.artisynth.org}).
This platform comes with tools to create physiologically accurate biomechanical models and is a good fit to bridge in between the RL research and the biomechanics.
ArtiSynth is developed in Java and is a cross-platform tool.

To transform ArtiSynth into an RL-friendly platform, a RESTful API was developed, as a plugin, to expose ArtiSynth's functionalities and facilitate development of reinforcement learning solutions (Figure~\ref{fig:artisynth-rl}).
This is a scalable server-client architecture which enables multiple ArtiSynth environments to run in parallel for sync/async RL training.
Moreover, given the fact that the modeling functionalities are exposed over a web based service conforming to the standard REST architecture, 
 researchers are given the flexibility to develop and train their RL solutions on any platform of choice and using any deep learning library.

To facilitate adoption of ArtiSynth-RL by the community, an OpenAI-Gym compatible class (\texttt{ArtiSynthBase}) was also developed in Python which translates utility commands (\emph{e.g.} reset, space sizes, etc.) into REST calls (\emph{i.e.} GET, POST). 
As a result, new environments can simply inherit from \texttt{ArtiSynthBase} and implement custom functionalities such as environment-specific reward functions.
The ArtiSynth plugins and the Gym environments are publicly released here: \url{https://github.com/amir-abdi/artisynth-rl}

\section{Experiments}

The reinforcement learning algorithm is implemented in Python using the PyTorch library.
The proximal policy optimization algorithm is  based on  OpenAI's open-source Baselines in TensorFlow~\cite{baselines},  extended by Kostrikov~\cite{pytorchrl}. 
The musculoskeletal model  is implemented  in the ArtiSynth platform.
The source code for the algorithms, RL plugins for the simulation platform, biomechanical model of spine, and run-scripts to reproduce the results reported in this paper can be found \href{https://github.com/amir-abdi/artisynth-rl}{here}.

\subsection{Musculoskeletal Model}
\label{sec:musc_model}

\begin{figure}[t]
\centering
\includegraphics[width=0.6\textwidth, trim=8 8 8 8, clip]{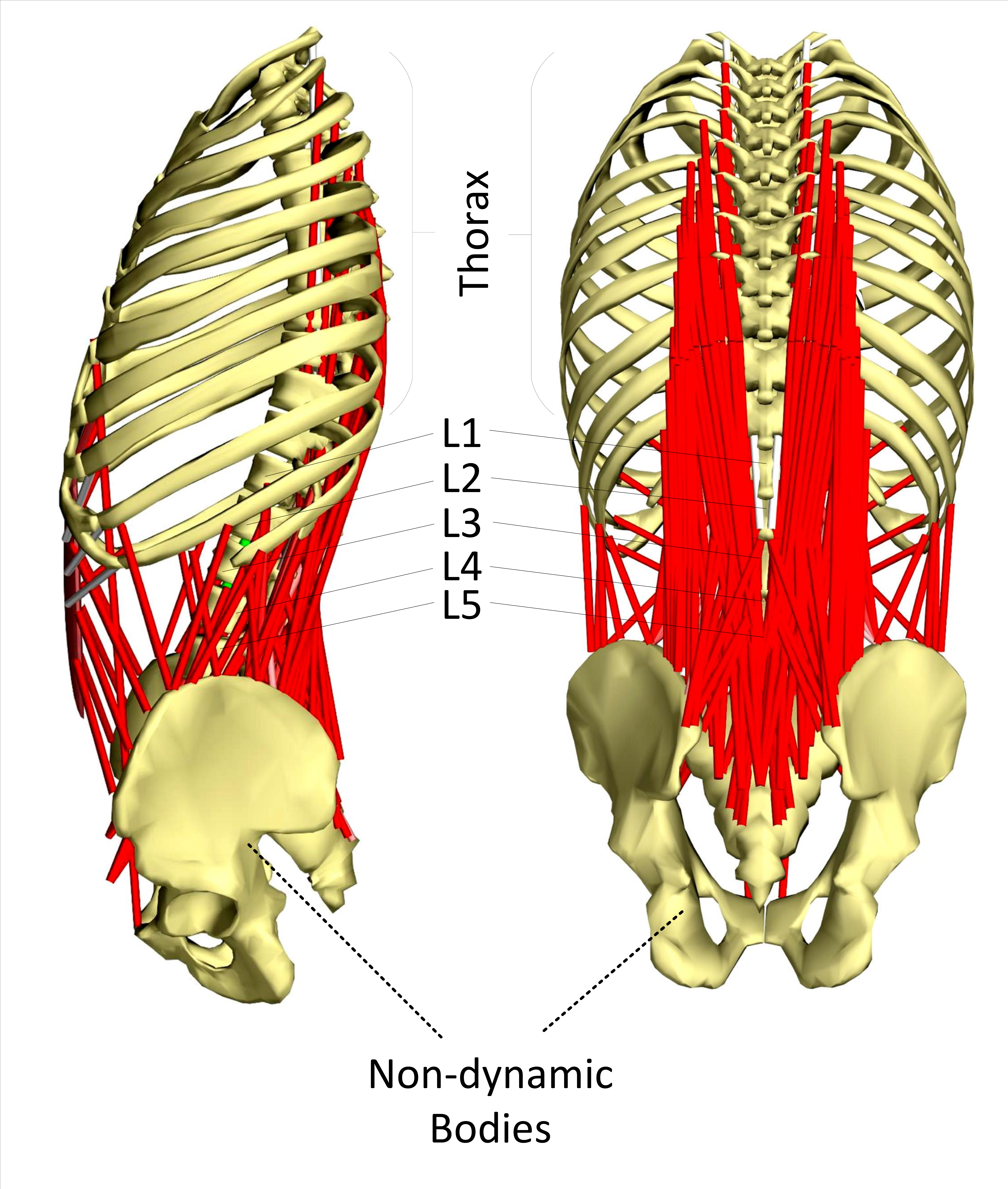}
\caption{The sagittal (right) and posterior (left) views of the musculoskeletal model of the lumbar spine, developed in the open-source platform of ArtiSynth.}
\label{fig:biomodel}
\end{figure}

This model is based on one of the most physiologically detailed musculoskeletal models of the lumbar spine which was originally developed in OpenSim~\cite{malakoutian2016musculoskeletal}.
The mechanical behaviors of the lumbar spinal units are modeled as a diagonal $6 \times 6$ stiffness matrix based on experimental data reported in the  literature. 
In all the experiments, both for the RL policy tracking and the FDAT, the  timestep of the simulation was set to 10 milliseconds.

This biomechanical system consists of five dynamic bodies, namely, the lumbar vertebrae (abbreviated as L1-L5), each with  6 degrees of motion (Figure~\ref{fig:biomodel}).
The functional spinal units  are simulated with frame springs,
which are six dimensional springs that generate restoring forces and moments between the connected bodies.
The frame springs are mass-less, thus, the forces and moments on their two ends are always equal and in opposite directions.

The topmost lumbar vertebra (L1) is fused to the thoracic vertebrae and the rib-cage.
The entire lumbar spine system rests on the non-dynamic pelvis and sacrum rigid bodies.
Since the lower-end of the model is static,
the orientation of the rib-cage suffices to describe its state in this constrained 3D space.
Consequently, the state of the entire system  includes the orientations of the five lumbar vertebrae (\emph{i.e.} L1 to L5).

Muscles are modeled as full Hill-type musculotendon actuators with tendon ratios and  pennation angles taken into account. 
Active properties of the muscles, including the active and passive force-length and force-velocity curves, were set based on the study by Millard~\cite{Millard2013}.
The physiologically accurate modeling of muscles adds a layer of complexity to the simulation due to the non-linear relationship between the neural excitations and the muscle force. 
The model consisted of 10 muscle groups, with a total of 210 muscle fascicles, each with a separate actuator (exciter). 
As a result, the model contained 210 actuated degrees of freedom.


The weight of the head is set at the physiological center of mass of the head and neck.
The effect of intra-abdominal pressure is modeled as an upward force at the center of the thorax.
Arms are considered to be symmetric and their mass and inertia are included in the body of the thorax.

\subsection{Training Strategies}

The RL policy is trained under a synchronous distributed training.
In each round of training, workers collect samples from parallel simulated environments. 
The training process waits for all workers to finish sample collection, merges rollouts, and updates the global policy networks according to equations ~\ref{eq:finalppo} and ~\ref{eq:trick_fixed}.

The state of the biomechanical model (Section~\ref{sec:musc_model}) is defined as the orientations  of the 5  rigid bodies (L1, L2, L3, L4, and L5) which are expressed as unit rotation quaternions.
Note that the lower end of the model is fixed (Figure~\ref{fig:biomodel}) and the state can be uniquely defined via the orientation; therefore, locations of the rigid bodies are left out of the equation. 
Moreover, the lumbar spine model is vertically super stiff to the extent that it can be realized as a multi-joint inverted pendulum with load.
Thus, slight vertical changes in the target positions  results in dramatic changes in muscle activations. 

The training  is designed in an episodic fashion.
At the beginning of each episode, 
the desired orientation of the thorax, and consequently the L1, are randomly set to a value within their feasible domain of rotation.
The desired orientations of L2, L3, L4 and L5 vertebrae are then set to $4/5$, $3/5$, $2/5$ and $1/5$ of L1's rotation, respectively.
This random setting of the desired lumbar vertebrae orientations results in a smooth bending of lumbar spine.

The agent would then explore the environment to reach the desired target orientations for the collective of lumbar vertebrae.
The agent reaches the terminal state when 
the squared second norm of differences between the orientations of the five lumbar vertebrae is less than the pre-defined threshold, $\delta$. 
The environment resets at the terminal state.
If the agent fails to reach the terminal state after $k$ steps, the environment resets.
After each reset, the thorax is randomly moved to a new orientation.

\subsubsection{Training Hyper-parameters}

Action space distributions are defined as diagonal Gaussians with standard deviations initialized as $exp(-1)$. This is approximately equal to one-third of the muscle excitation range of [0 - 1] which ensures sufficient initial exploration.

The actor and critic networks consist of four and two fully-connected (FC) layers, respectively. 
The number of neurons in the actor network start with 32 and double  every layer.
The breadths of the critic network layers are set to 64.
All FC layers are followed by a $tanh$ activation function.
Outputs of the actor network, \emph{i.e.}, action features, are processed by an FC layer to calculate the mean parameters of the multi-dimensional Gaussian distribution of the action space.

Due to the high actuated degrees of freedom, the  entropy regularization coefficient, $c_s$, was mitigated to 0.001. 
The value loss coefficient, $c_v$, was set to 0.5.
The clipping term ($\epsilon$) and the learning rate were initialized at 0.2 and 0.007.
Once the training converges to an approximate solution, the clipping term and learning rate were scheduled for linear decay. 
Parameters of the actor-critic networks were optimized using Adam with a mini-batch size of 16 and 32 steps per environment per policy update.
The $\tau$ parameter of Generalized Advantage Estimation was set to $0.95$ and $\gamma$ was set to $0.99$.
The number of steps before resetting the episode was $k=30$.



The three coefficients in Eq.~\ref{eq_total_qp} and Eq.~\ref{eq:reward} require manual tuning to achieve desirable results. 
We ran a linear grid search in the coefficients' spaces of $w_u$, $w_d$, and $w_r$ to find a set of suitable configurations that perform reasonably well on both approaches, \emph{i.e.} RL
and FDAT.
For the results of the two approaches to be comparable, we  used the same coefficients in the reward function of the RL approach as those in the objective function of the FDAT optimizer.
The coefficients used in the reported results are as follows: 
$w_u=1.0,~ w_r=0.01,~ w_d=0.001$.





\section{Results and Discussion}

\begin{table}
\centering
\begin{tabular}{lccccc}  
\toprule
Method & $error_c$  &  $error_s$  & $error_t$ & $error_e$ & Time\\
\midrule
\textbf{FDAT}  & 0.0008 & 0.0001 &  0.0028 $\pm$ 0.0148 &  1.4854 $\pm$ 2.1091 & 10.9\\
\textbf{RL}  & 6.8254 & 0.0011 &  0.0 $\pm$ 0.0 &  9.8065 $\pm$ 2.1651 & 1.0\\
\bottomrule
\end{tabular}
\caption{Comparison of the performance of the two methods across the four metrics.
The correctness error ($error_c$) and stability error ($error_s$) are represented in degrees. Since  muscle excitation is a relative quantity, the temporal consistency error ($error_t$) and efficiency error ($error_e$) are unit-less.
The run-time of the two solutions are expressed in relative units to highlight that the FDAT approach is $10\times$ more computationally intensive compared to the neural policy of the RL approach.
}
\label{table:results}
\end{table}

The learned RL policy is evaluated in terms of stability, temporal consistency, energy efficiency, and correctness.
The results of these evaluations are compared with the numerical solution of FDAT.
The models were evaluated over 60 random, but consistent,  target-reaching experiments, with 100 samples collected from each trial.
The results of these evaluations are reported in Table~\ref{table:results}.

\paragraph{Stability ($error_s$)}
is defined as the lack of motion jitter,  during the target reaching task and in static positions. 
The stability error is calculated as the standard deviation of distances to the desired position.

\paragraph{Correctness ($error_c$)} is defined as the Euclidean distance between the model's position and the desired position. 

\paragraph{Temporal consistency  ($error_t$)} 
is the average difference between consecutive muscle excitations during an episode of motion, which is calculated as  $\mathrm{E}\Big[\norm{\textbf{a}_t-\textbf{a}_{t-1}}\Big]$.

\paragraph{Efficiency ($error_e$)} is evaluated with respect to the energy consumption of the  biomechanical model.
Energy-efficiency is calculated as the average norm of muscle excitations during an episode of motion, calculated as $\mathrm{E} \Big[\norm{\textbf{a}_t}\Big]$.

Like any other numerical solver, the iterative nature of the Dantzig-Cottle PPM method can generate different results in between consecutive timesteps even if the state of the environment has not changed. 
Consequently, the model can bounce for a while until eventually becoming stable. Increasing the damping and regularization coefficients ($w_d$ and $w_r$) makes this problem even worse and increases  the stability error ($error_s$).
As demonstrated in Table~\ref{table:results}, the reinforcement learning policy results in a more temporally consistent solution with the same level of stability compared to the FDAT approach.

We also conducted a side-by-side run-time analysis of the two solutions on the same machine and under the same conditions.
According to this  analysis, the neural network based policy performed 10.9 times faster than the quadratic solver of the FDAT (Table~\ref{table:results}).
Solving an LCP  with an arbitrary square matrix is considered NP-complete~\cite{LCPNPComplete}; however, it can be shown that Dantzig and Cottle's principal pivoting method can be solved in $2^n-1$ steps for a positive semi-definite matrix of size $n$. 
The same stands true for similar solutions such as  Van der Heyden's variable dimension algorithm~\cite{complexityDantzig}.
Only under hard assumptions, which are improbable in computational mechanics, polynomial solutions can be sought~\cite{polyLCP}.
As a result, in mechanical systems with hundreds of degrees of freedom, the computational complexity of such solvers is more than what a real-time forward dynamics simulation can handle. 
In contrast, an RL policy (actor), parameterized by a deep neural network with a multitude of layers, can calculate actions with a single feed-forward pass in a substantially shorter time.

The  biomechanical system in question can be realized as a multi-joint inverted pendulum with load,  and with an added level of complexity in  motor control.
Moreover, the agent dealt with  a partially observable environment where the  joint angles were not disclosed. 
As a result of these twists, the trained RL actor failed to reach the desired location  in some of the random test cases, which were included in the performance analysis, resulting in a higher $error_c$ compared to the static solver.
Interestingly, all the mentioned inaccessible locations where posteriorly located with respect to the thorax. 
Therefore, it is our understanding that the extension position  of the lumbar spine (upper body extending backwards) is an intricate task to master and requires a more thorough tuning.

Based on the reported results, and while considering the limitations of the current study, it was found that penalizing higher muscle excitations in reinforcement training is not as effective as constraining excitations in a static numerical solver.
This is to the extent that the smallest of regularization coefficients ($w_r$) are quite effective in the FDAT solver, but the same does not stand true in the implemented RL-PPO approach.


\section{Conclusion and Future Work}

There is an ongoing  effort to model the neural control of musculoskeletal systems~\cite{MLinBio}. 
In this paper, we put forward a framework to track movements of a physiologically accurate biomechanical model using a parameterized policy trained with the PPO RL algorithm. 
The performance of the trained RL policy was compared with the  well-known static optimization method of forward dynamics assisted tracking (FDAT). 

Given the high-dimensional continuous action space and the partially observable environment, the convergence of the RL model was found to be challenging.
The results point to the fact that the RL approach is stable and temporally consistent in terms of muscle excitations. 
Moreover, the RL solution tends to run  $10\times$ faster compared to the static solver of FDAT, making it the go-to-option for real-time inverse tracking.
However, the agent failed to reach the desired target orientations, mostly in the posterior (extension) body postures. This issue might be mitigated by making the environment fully observable as well as coupling muscle excitations to reduce the action space size.
 
Future research on more physiologically detailed models and with more comprehensive tests can help advance our understanding of the neural control of human locomotion. 
To pave the way, the ArtiSynth mechanical platform was reshaped into an RL-ready environment through a web based API, which can be integrated with familiar reinforcement learning frameworks.

\paragraph{Acknowledgement} 
 We would like to thank John Lloyd, Ian Stavness, and Antonio Sanchez for their priceless guidance. 
 This research was conducted thanks to the Vanier Scholarship by NSERC Canada awarded to the first author.


\bibliography{neurips_2019}
\bibliographystyle{plain}

\end{document}